\documentstyle[12pt,epsfig]{article}
\def\beq{\begin{equation}}
\def\eeq{\end{equation}}
\def\bea{\begin{eqnarray}}
\def\eea{\end{eqnarray}}

\begin{document}

\rightline{DESY 98-040}
\rightline{hep-ph/9803495}
\rightline{March 1998}
\vspace{1.5cm}

\begin{center}
{\large \bf NUCLEAR EFFECTS IN QCD}\footnote{Talk given
at the {\it 5th International Workshop on Hard Probes in Nuclear
Collisions} (Lisbon, Portugal,
September 7th-18th 1997).}
\vspace{1cm}

N. Armesto\\
{\it II. Institut f\"ur Theoretische Physik, Universit\"at
Hamburg,\\
Luruper Chaussee 149, D-22761 Hamburg, Germany}\\
\end{center}
\vspace{4.0cm}

Production of Drell-Yan
pairs and open and hidden heavy flavor on
nuclear targets is
examined within perturbative QCD.
The effects of modifications of
nucleon
structure functions inside the nuclear medium are considered.
Besides,
nuclear dependence of charmonium and bottonium absorption is studied in
the framework of the Glauber-Gribov model.
The low energy limit of this approach
recovers the probabilistic formula usually employed
for charmonium and bottonium
suppression in nuclear collisions.

\vfill

\newpage

\section{Introduction}\label{sec1}

Drell-Yan ($DY$) and open ($Q\overline{Q}$) and hidden ($\Psi$)
heavy flavor production in hadron
collisions are usually
studied in the framework of perturbative QCD \cite{mangano,braaten}. The
extension to nuclear projectiles and targets is straightforward,
provided factorization between partonic densities and parton-parton
cross sections is assumed:
\beq
\sigma_{AB}\equiv
\sigma^{AB\rightarrow h\overline{h}} = \sum_{a,b} \int_{x_{a0}}^1
dx_a \int_{x_{b0}}^1 dx_b
\  f_{a/A}(x_a,\mu^2)
f_{b/B}(x_b,\mu^2)\ \hat{\sigma}^{ab\rightarrow h\overline{h}}
(\hat{s},m_h,\mu^2). \label{eq1}
\eeq
In this equation the summation runs over all partons in the projectile
and target, $f_{c/C}$ is the density of parton $c$ in hadron or nucleus
$C$, $\hat{\sigma}^{ab\rightarrow h\overline{h}}$ is the parton-parton
cross section and factorization and renormalization scales are taken to
be equal;
$m_h =m_Q$ ($M_{l^+l^-}/2$) for heavy flavor (Drell-Yan) production,
$x_{a0}=4m_h^2/s$, $x_{b0}=4m_h^2/(sx_a)$,
$\hat{s}=sx_a x_b$ and $\mu^2 \sim m_h^2$. The leading order (LO) relation
$x_F=x_a-x_b$ will be used.

In this context,
some differences between $A$, $B$ being hadrons or nuclei are the
following:
\begin{itemize}
\item The influence of the nuclear medium on nucleon
structure functions or partonic densities (see for example \cite{emcr}),
expressed as the ratio:
\beq
R_{a/A}(x,\mu^2)=\frac{2\ f_{a/A}(x,\mu^2)}{A\ f_{a/D}(x,\mu^2)}\neq 1.
\label{eq2}
\eeq
If we parametrize cross sections on nuclei as
$\sigma_{AB}(x_F) =\sigma_{pp}\ (AB)^{\alpha(x_F)}$, this effect makes
$\alpha(x_F)\neq 1$ for all processes,
even if isospin effects (i.e. the difference
between neutron and proton parton densities) are corrected.
\item The scattering of the produced
heavy system (absorption
by nuclear matter, rescattering with hadronic co-movers and/or deconfinement,
see \cite{ollitrault}), which affects $\Psi$ production, making
$\alpha^{\Psi}(x_F)<\alpha^{DY}(x_F)$.
\item The elastic scattering of initial partons, which is the accepted
explanation of the
$p_T$-broadening
($\langle p_T^2
\rangle_{AB}$ greater than $\langle p_T^2
\rangle_{pp}$ for $DY$ and $\Psi$ production),
being
the difference proportional
to $A^{1/3}$ \cite{ollitrault,sterman,gv}.
\item The energy loss of fast partons inside the nuclear medium (the
so-called jet quenching), proportional to $A^{2/3}$ and different
for cold and hot
nuclear matter \cite{sterman,baier}. It may affect the yield of charm
and bottom at high energies \cite{vogt}.
\end{itemize}

In this contribution we will examine the two first aspects. It is
organized as follows: In Section \ref{sec2}
nuclear structure
functions will be briefly discussed, and the parametrizations
used in our calculations
and results for Drell-Yan, open heavy flavor and charmonium production
on nuclear targets
will be presented. In Section \ref{sec3} nuclear
absorption of states with hidden heavy
flavor will be studied. Finally, in Section \ref{sec4} some conclusions will be
presented.

\section{Hard processes on nuclear targets}\label{sec2}

Since the experiments of the European Muon Collaboration \cite{emc}, the
modification of nucleon structure functions inside the nuclear medium has
been studied by several experiments \cite{emcr}. While the dependence of
$R_{a/A}(x,\mu^2)$ on
$\mu^2$ is very small, four regions in $x$ can be described: $R_{a/A}>1$
for $0.8 < x < A$ (Fermi motion and cumulative regions);
$R_{a/A}<1$ for $0.3 < x < 0.8$ (the original EMC effect);
$R_{a/A}>1$ for $0.1<
x<0.3$ (antishadowing region); and $R_{a/A}<1$ for $x<0.1$ (shadowing
region).

We will use the parametrization of Ref. \cite{esk93}\footnote{An update of
this parametrization can be found in \protect{\cite{esk98}}; among
other modifications, modern sets of nucleon structure functions have
been used. Other proposal in the same spirit can be found in
\protect{\cite{ind}}.} for $Au$, which was designed to
describe the four regions in $x$. This approach follows the conventional
techniques for global fits to produce nucleon parton distributions: the
ratio $R_{a/A}$ is parametrized at some low virtuality (4 GeV$^2$) and
then evolved to higher $\mu^2$ using evolution equations modified for
the nuclear case \cite{mq}. All parameters are fixed from a comparison
to nuclear structure function ratios over deuterium and $DY$ data, using
baryon number and momentum sum rules and a $SU(3)$ symmetric sea.
This parametrization has a lower limit in $x=10^{-3}$
($\equiv \sqrt{s}\simeq 100$ GeV for charm at $x_F=0$,
not enough for predictions for RHIC and LHC
or for
high $x_F=x_a-x_b$), so at smaller $x$ we have taken two
alternatives: either frozen or a linear-log extrapolation in the form
$x^\beta$. The parametrization for two different virtualities can be
seen in Fig. 1 \cite{noso}.
\begin{figure}[hbt]
\caption{Parton densities in $Au$ for $\mu^2=5$ and 25 GeV$^2$: valence
quarks (dashed curve), sea quarks (solid curves) and gluons
(dashed-dotted curves); dotted curves are the results of linear-log
extrapolations with $\beta=$ 0.096 and 0.040 at $\mu^2=5$ and 25 GeV$^2$
respectively.}
\epsfig{file=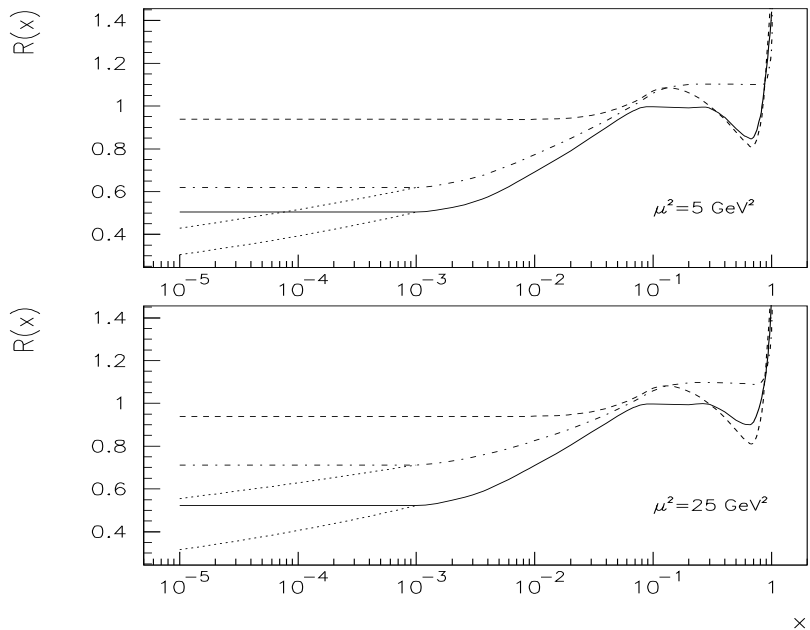,width=15.5cm,height=13cm,
angle=0,bbllx=0pt,bblly=0pt,
bburx=600pt,bbury=650pt}
\end{figure}
\vskip -3 cm
\noindent {\large \bf 2.1 Open heavy flavor and Drell-Yan}

\noindent
Parametrizing $\sigma_{pA}(x_F) =\sigma_{pp}\ A^{\alpha(x_F)}$,
in Figs. 2-5
results \cite{noso} for $\alpha(x_F)$ in $DY$, $c$ and $b$ production
in $pAu$ collisions are presented,
using GRV HO \cite{grv94} and MRS A \cite{mrs}
nucleon structure functions.
The $\hat{\sigma}^{ab\rightarrow h\overline{h}}
(\hat{s},m_h,\mu^2)$ are taken at next-to-leading order in the
$\overline{\mbox{MS}}$ renormalization scheme and the
following parameters are used:

\noindent i) For open heavy flavor \cite{nde}:
$m_c=1.5$ GeV and $\mu^2=4$ GeV$^2$
for charm, and $m_b=5$ GeV and $\mu^2=m_b^2$ for bottom.

\noindent ii) For Drell-Yan \cite{fp}:
$\mu^2=M^2_{l^+l^-}$.

As stated above, $x_F=x_a-x_b$ (LO relation), $x_ax_b s\geq 4m_h^2$ and the
main contribution to the integrals comes from
$ x_ax_b s=4m_h^2$, $x_{a/b}=[\sqrt{(16m_h^2/s)+x_F^2}\pm x_F]
/2$; this means
$x_{a/b}\simeq 3\cdot 10^{-4}$ for charm at
$\sqrt{s}=10$ TeV and $x_F=0$.
\begin{figure}[hbt]
\caption{Energy dependence of $\alpha$ for charm and beauty
production in $pAu$ collisions for GRV HO (solid and dashed
curves) and
MRS A (dotted and dashed-dotted curves) structure
functions and using extrapolated (dashed and dashed-dotted curves) and
frozen at
$x= 10^{-3}$ (solid and dotted curves) ratios of parton distributions.}
\epsfig{file=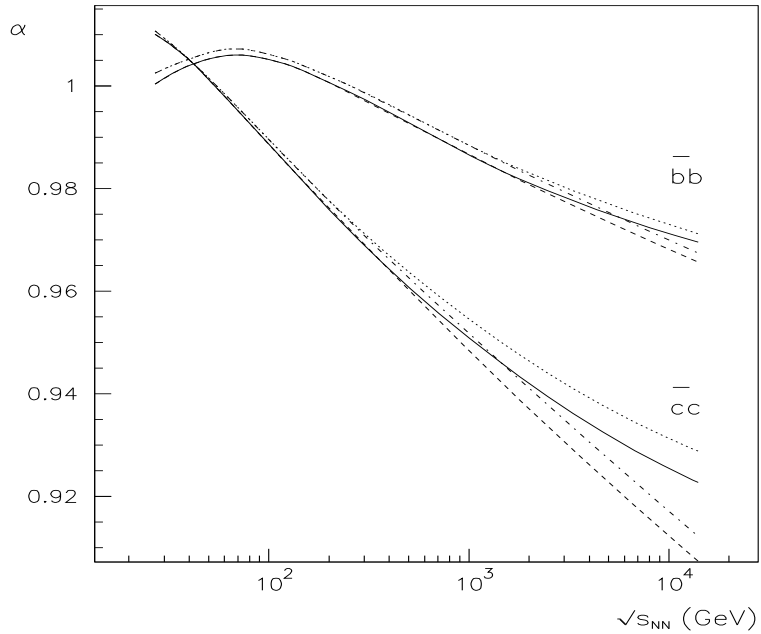,width=15.5cm,height=13cm,
angle=0,bbllx=0pt,bblly=0pt,
bburx=600pt,bbury=650pt}
\end{figure}
\vskip -3 cm

It can be seen in these Figures that the influence of the chosen Set of nuclear
parton densities is almost negligible except for the highest energies
and $x_F$.
Besides, some difference appears between frozen and extrapolated ratios;
as expected,
this difference is larger at higher energies or $x_F$ and for charm or
low dilepton masses. Moreover, in the $x_F$ distributions all regions in
nuclear structure functions can be seen from negative to positive $x_F$
(corresponding to decreasing $x$). Related results can be found in Refs.
\cite{mw,rvogt}.
\begin{figure}[hbt]
\caption{$x_F$ dependence of $\alpha$ for charm
and beauty production in $pAu$ collisions at $\protect{\sqrt{s_{NN}}}$
= 39
GeV
(upper figure) and 1800 GeV (lower figure)
for GRV HO and MRS A structure functions and using
extrapolated and frozen at $x = 10^{-3}$
ratios of parton distributions
(with the same conventions as in Fig. 2).}
\epsfig{file=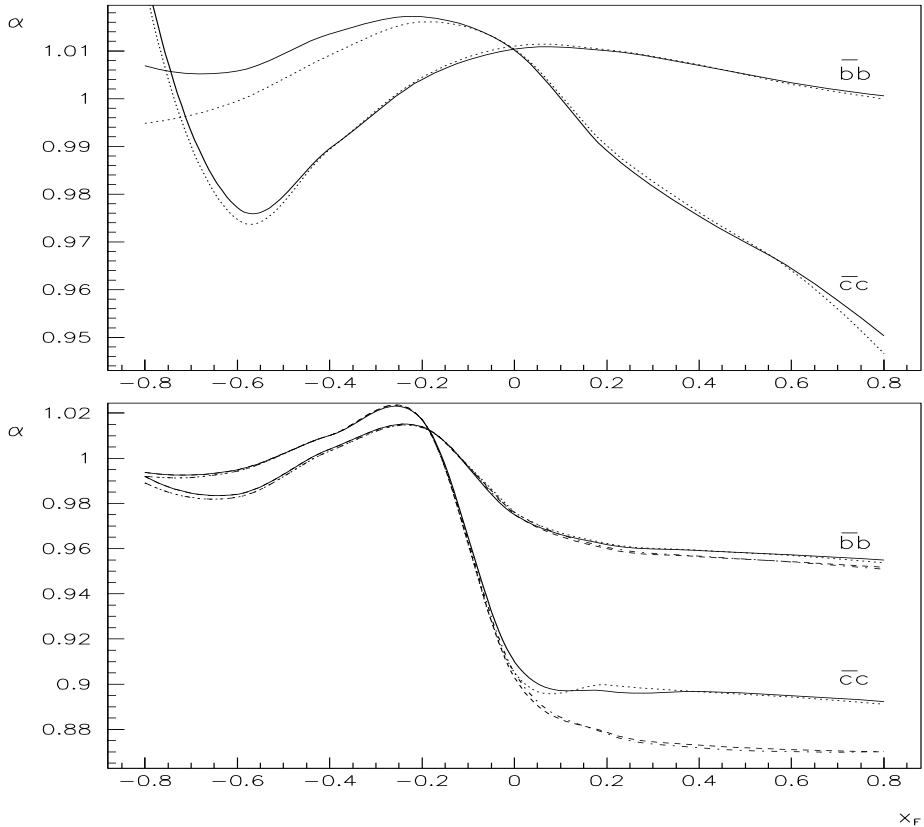,width=15.5cm,height=13cm,
angle=0,bbllx=0pt,bblly=0pt,
bburx=600pt,bbury=750pt}
\end{figure}
\begin{figure}[hbt]
\caption{Mass (upper figure)
and energy (lower figure)
dependence of $\alpha$ for
Drell-Yan pair production in $pAu$ collisions for GRV HO and MRS A
structure functions and using extrapolated and
frozen at $x = 10^{-3}$ ratios of parton distributions
(with the same conventions as in Fig. 2). In the lower figure
$M^2$ is in GeV$^2$.}
\epsfig{file=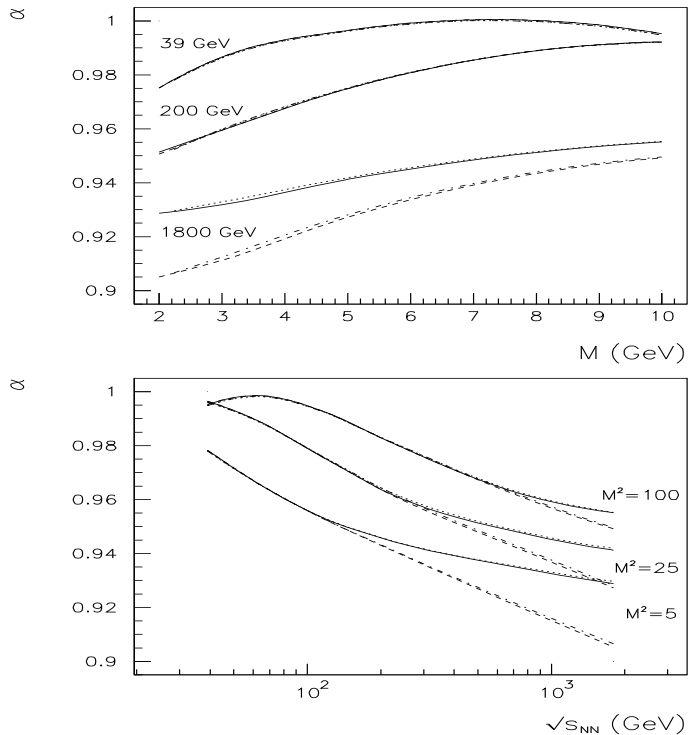,width=15.5cm,height=13cm,
angle=0,bbllx=0pt,bblly=0pt,
bburx=600pt,bbury=750pt}
\end{figure}
\begin{figure}[hbt]
\caption{$x_F$ dependence of $\alpha$ values for heavy
lepton pair production in $pAu$ collisions at $\protect{\sqrt{s_{NN}}}$
= 39
GeV
(upper curves in each figure) and 1800 GeV (lower curves in each figure)
and different masses
for GRV HO and MRS A structure functions and using
extrapolated and frozen at  $x = 10^{-3}$
ratios of parton distributions
(with the same conventions as in Fig. 2). Note that at $\sqrt{s_{NN}}$ =
39 GeV
all curves coincide.}
\epsfig{file=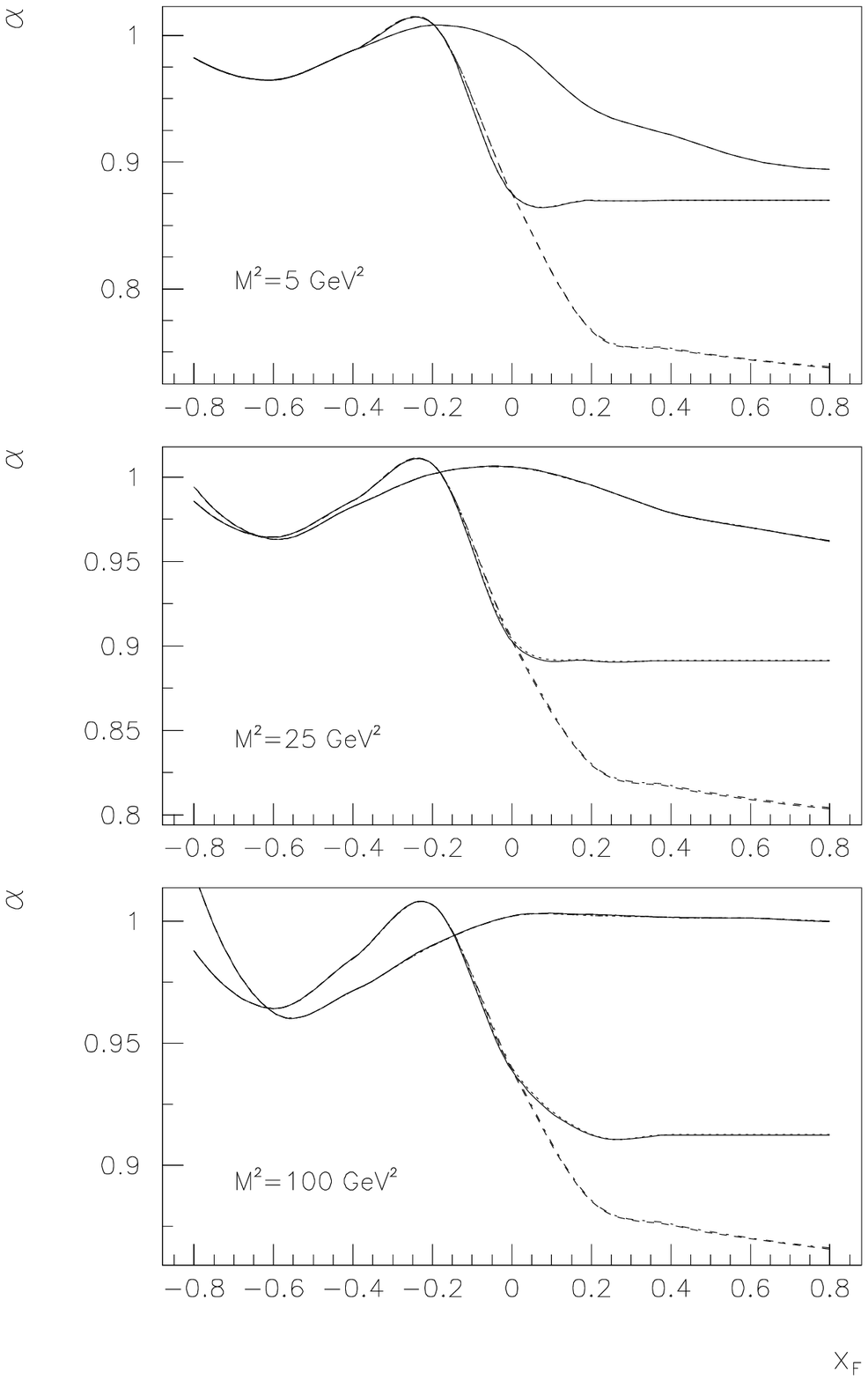,width=15.5cm,height=11.5cm,
angle=0,bbllx=0pt,bblly=0pt,
bburx=600pt,bbury=750pt}
\end{figure}

\vskip 2 cm
\noindent {\large \bf 2.2 Hidden heavy flavor}

We will concentrate on charmonium production on nuclear targets.
Now the formation of the final resonance has to be considered. Usually two
models are used to study charmonium production:

\noindent i) The Color Evaporation Model (CEM) \cite{cem} considers
that all color dynamics is contained in the kinematical restriction to
Eq. (\ref{eq1}):
$4m_c^2 \leq \hat{s}=x_ax_bs\leq 4m_D^2$. The projection on different
charmonium states is taken into account by numerical coefficients,
$\sigma^{AB\rightarrow \Psi} = F_\Psi \ \sigma^{AB\rightarrow
c\overline{c}} (4m_c^2 \leq \hat{s}\leq 4m_D^2)$,
which are universal in this approach.

\noindent ii) The Factorization Approach (FA), which contains both the
Color Singlet Model (CSM) and Color Octet Model (COM), is based on
non-relativistic QCD (NRQCD) \cite{braaten,nrqcd}. In this model the
parton-parton cross section for production of a charmonium state $\Psi$
is:
\beq
\label{eq3}
\hat{\sigma}^{ab\rightarrow
\Psi}
(\hat{s},m_c,\mu^2)=\sum_{[n]} C^{ab}_{c\overline{c}[n]}\langle
O^\Psi_{[n]}\rangle,
\eeq
with
$C^{ab}_{c\overline{c}[n]}$ the short distance coefficients for the hard
subprocess $ab\rightarrow c\overline{c}[n]$ ($[n]$ is the color
configuration of the $c\overline{c}$ pair),
computable as series in $\alpha_s$, and
$\langle O^\Psi_{[n]}\rangle$
the long distance matrix elements taking into account the
ha\-dro\-ni\-za\-tion
$c\overline{c}[n]\rightarrow \Psi$, which
can be classified in powers of the relative velocity $v$ of $Q$ and
$\overline{Q}$ and are obtained from a fit to data \cite{BR,TV,GS}.
We will use the fit
to fixed target data obtained in Ref. \cite{BR}.

While in CEM different resonances have the same energy and $x_F$
behavior, in FA we
consider different contributions and take into account
the most important decays from
charmonium states into $J/\psi$: 
\beq
\label{eq4}
\sigma_{J/\psi}^{tot}= \sigma_{J/\psi}^{dir} + B(\psi^\prime
\longrightarrow J/\psi)\ \sigma_{\psi^\prime} 
+\sum_{J=0,1,2} B(\chi_{cJ}\longrightarrow J/\psi)\
\sigma_{\chi_{cJ}},
\eeq
being $\sigma_{J/\psi}^{dir}$ the direct $J/\psi$ production, i.e. not
coming
from decays, and $B(H\to J/\psi)$ the branching ratios for particle $H$
to
decay into a $J/\psi$. Experimentally it is found that
direct $J/\psi$ contribution is about 60 \% of the
total $J/\psi$ cross section, $\psi'$ decay gives less than 10 \% and all
$\chi_{cJ}$
contribute with more than 30 \%.
Decays into $\psi'$ are not important, and then $\psi'$ production is
dominated by direct production.

One important point is the contribution of the different color states to
the
production of these particles, i.e. the color content of the
pre-resonant state.
In fact, the FA gives that
direct $J/\psi$ production is almost completely produced in color octet
state,
$\psi'$ is also
predominantly (about 90 \% or more) produced in color octet, and
the main contribution of the $\chi_{cJ}$ states to $J/\psi$ comes from
color
singlet matrix elements. Then, a separate study of the effect of nuclear
structure functions on
different
particles and color states is possible in this approach.
As we will see, the color octet and color singlet
contributions to the production of charmonium have different suppression.

In Figs. 6 and 7 we present results for energy and $x_F$ dependence of
charmonium production \cite{pss}.
Computations have been done taking $m_c=1.5$ GeV and $\mu=2m_c$, and using
CTEQ3L \cite{cteq} and GRV HO \cite{grv94} nucleon parton distributions
in FA and CEM respectively. Different color contributions and charmonium
states are taken into
account defining
\beq
\label{eq5}
\alpha_i^\Psi(x_F)= \frac{\ln{[\sigma^{\Psi,i}_{AB}(x_F)/
\sigma^{\Psi,i}_{pp}(x_F)]}}{\ln{A}},
\eeq
for $i=$ CSM, COM, CSM+COM=FA, CEM, and
$\Psi=$ total $J/\psi$, direct $J/\psi$, $\psi^\prime$,
$\sum B(\chi_{cJ}\longrightarrow J/\psi)\ \sigma_{\chi_{cJ}}$.
It can be observed that both FA and CEM give very similar results except
for $\chi$'s at
very high energies or $x_F$. The behavior of the different color
contributions (and hence of the $S$ wave and $P$ wave states) is
also different at high energies or $x_F$.
\begin{figure}[hbt]
\caption{Center of mass energy dependence of $\alpha$ for $pAu$
collisions.
Different lines are:
FA total contribution (solid), singlet contribution (dotted), octet
contribution (dashed) and CEM (dashed-dotted).}
\epsfig{file=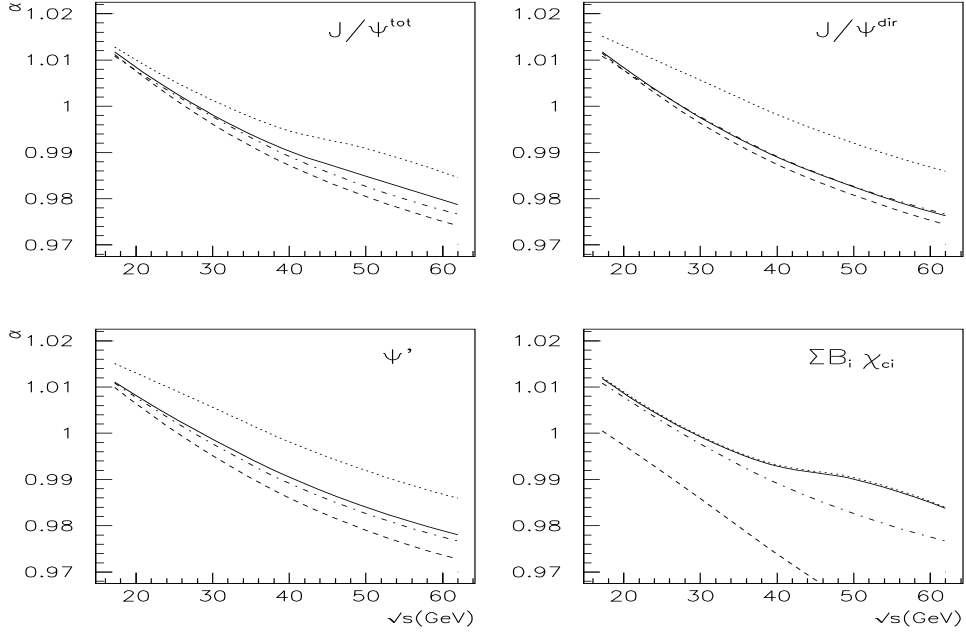,width=15.5cm,height=13cm,
angle=0,bbllx=0pt,bblly=0pt,
bburx=600pt,bbury=750pt}
\end{figure}
\begin{figure}[hbt]
\caption{$x_F$ dependence of $\alpha$ for $\protect{\sqrt{s}}=39$ GeV
$pAu$ collisions with the same
conventions for lines as in Fig. 6.}
\epsfig{file=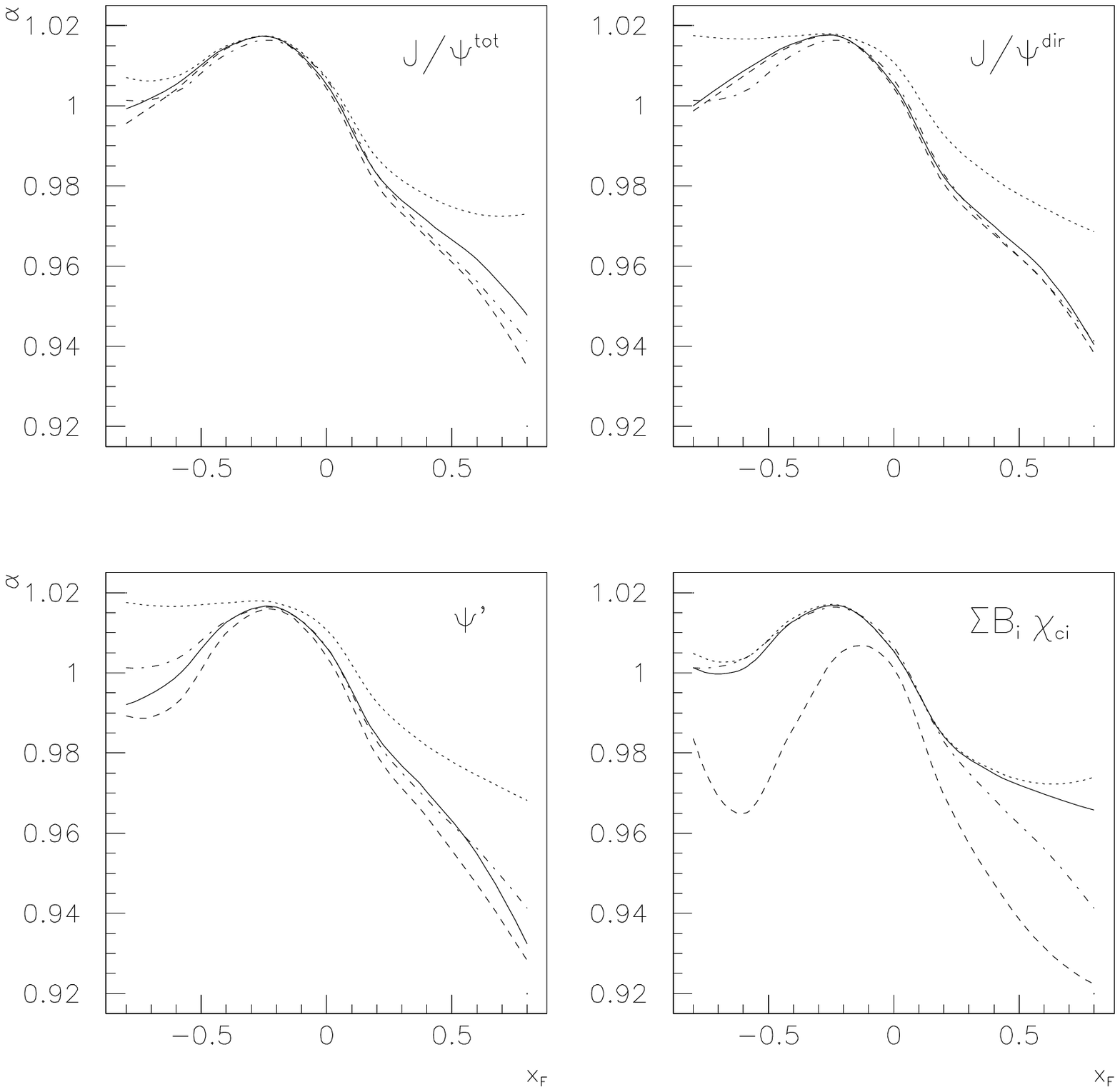,width=15.5cm,height=13cm,
angle=0,bbllx=0pt,bblly=0pt,
bburx=600pt,bbury=750pt}
\end{figure}

\section{Nuclear absorption}\label{sec3}

Nuclear absorption of the pre-resonant
$c\overline{c}$ state in its path through nuclear matter is usually
taken into account in $pA$ collisions,
at fixed impact parameter $b$, by two formulae (which
neglect nuclear effects on structure functions, assumed very small at
fixed target energies in the central rapidity region):

\noindent i) Taking into account the path $L(b)$ across the nucleus
\cite{ger}, Fig.
8:
\beq
\label{eq6}
\sigma_{pA}^\Psi(b) \propto
\exp[-\rho_0 \sigma_{abs} L(b)],\ \  \rho_0\simeq 0.17\ \ 
\mbox{fm}^{-3}.
\eeq

\noindent ii) The Glauber probabilistic formula \cite{alfons}:
\bea
\sigma_{pA}^\Psi &=& \sigma_{pN}^\Psi \ A\int d^2b\ 
\int_{-\infty}^{+\infty} dz\
\rho(z,\vec{b})\ \exp{\left[
-\sigma_{abs}A\int_z^{+\infty}dz^\prime
\rho(z^\prime,\vec{b})\right]}\nonumber \\
&=&\frac{\sigma_{pN}^\Psi}{\sigma_{abs}}
\ \int d^2b\ \left[1-e^{-\sigma_{abs}AT_A(b)}\right];\label{eq7}
\eea
this formula can be easily understood (Fig. 8): the pre-resonant
$c\overline{c}$ state is created at some point $z$ in the nucleus (in an
amount proportional to the density of nuclear matter $\rho(z,\vec{b})$
at that point) and
is absorbed in its path through the nucleus from $z$ to $+\infty$ with
some absorption cross section\footnote{The meaning of this absorption cross
section is
not clear. In some proposals (e.g. \cite{Kh94}) it has been related to the
color structure of the pre-resonant state: absorption is much stronger
in octet than in singlet configuration. Observation of little
absorption for $\chi$'s (see Subsection 2.2)
and variation of absorption with $p_T$ of the produced resonance
in some defined way would support
this point of view and the validity of the NRQCD approach.}.
For open charm ($\sigma_{abs}=0$), $\sigma_{pA}=\sigma_{pN}
\ A$.
\begin{figure}[hbt]
\label{fig8}
\caption{Representation of the absorption mechanism at low energies in
the rest frame of the nucleus; $z$
is the creation point of the pre-resonant
$c\protect{\overline{c}}$ pair.}
\hspace{2cm}\epsfig{file=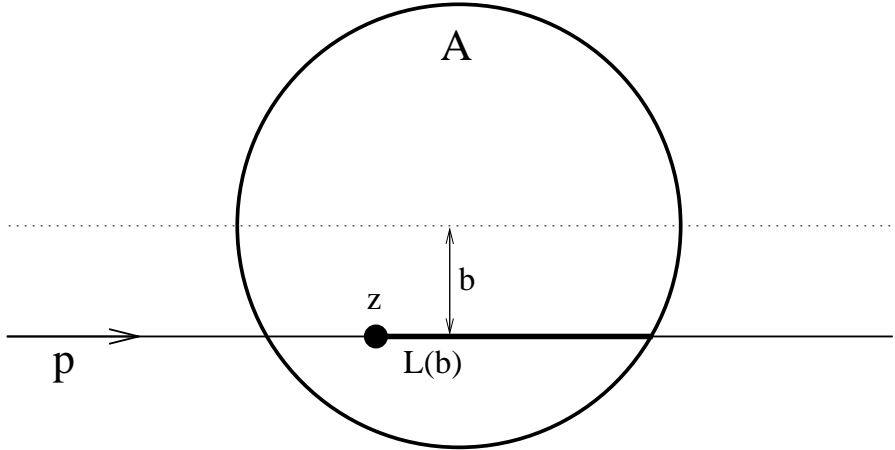,width=6cm,angle=270}
\end{figure}

Eqs. (\ref{eq6}) and (\ref{eq7}), with their longitudinal ordering, are
only valid in the low energy limit. 
A formula valid at all energies has been derived \cite{Br98}
in a fully relativistic Glauber-Gribov
approach, using finite energy Abramovsky-Gribov-Kancheli (AGK)
cutting rules \cite{Br94}. Considering $n$
nucleon-nucleon
interactions taken
in some arbitrary longitudinal ordering $z_1\leq z_2\leq\cdots
\leq z_n$, these rules imply (besides the usual AGK prescription
\cite{agk}) to change $T_A^n(b)$,
$T_A(b)=\int_{-\infty}^{+\infty}dz\ \rho(z,b)$, by
\beq
\label{eq8}
T_n^{(j)}(b)=n!\int_{-\infty}^{+\infty}dz_1
\int_{z_1}^{+\infty}dz_2\cdots \int_{z_{n-1}}^{+\infty}dz_n
\ \exp{[i\Delta (z_1-z_j)]}\prod_{i=1}^n \rho(z_i,b),
\eeq
for $j$ the first interaction either cut or to the right of the cut (see
Fig. 9)
and
\beq
\label{eq9}
\Delta=\frac{m_N M^2}{s x_a},\ M=2m_h\ \mbox{or}\ m_\Psi.
\eeq
This is nothing but the $t_{min}$ effect \cite{Bo93}, i.e. the nuclear
form factor suppresses interactions which require a minimum momentum
transfer
(e.g. those in which heavy flavor is produced).
\begin{figure}[hbt]
\label{fig9}
\caption{Position of the cut to the left of the $j$-th interaction (dotted
line) or cut in the $j$-th interaction (crossed solid line).}
\hspace{1cm}\epsfig{file=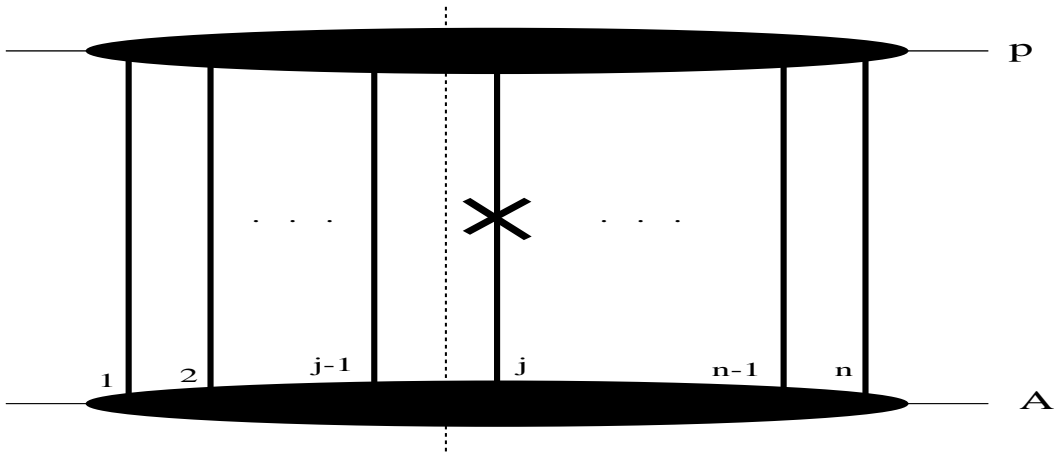,height=14cm,width=6cm,angle=270}
\end{figure}

Both an external contribution (corresponding to the heavy system being
produced in the interaction) and internal\footnote{Not to be confused
with intrinsic charm \cite{brodsky}, which can be important
at high $x$.} contributions
(corresponding to the heavy system already present
in the projectile or target, which turn out to be very small)
are included. Modifications
of the nucleon structure functions due to the nuclear medium naturally
appear in this framework, which is valid at not very high $x$. The results for
the low and high energy asymptotic limits are:

\noindent i) For low energies, $\Delta\longrightarrow \infty
\Longrightarrow$ only $j=1$ contributes, the shadowing of structure
functions disappears and we recover the probabilistic formula
(\ref{eq7}). To our knowledge it is the first time that this equation has
been derived in a fully relativistic approach.

\noindent ii) For $s\longrightarrow \infty$, $\Delta = 0$ and $T_n^{(j)}(b)
= T_A^n(b)$, the shadowing of the structure functions
factorizes and we get for the external part:
\bea
\sigma_{pA}^\Psi &=& \sigma_{pN}^\Psi\ \frac{2}{\sigma}\int d^2b
\ e^{-\tilde{\sigma}AT_A(b)/2}
\left[ 1-e^{-\sigma AT_A(b)/2}\right] \label{eq10} \\ &\longrightarrow&
\sigma_{pN}^\Psi\ A\int d^2b \ T_A(b)\ e^{-\tilde{\sigma}AT_A(b)/2}\ \ \
\ \mbox{(if no shadowing).}
\label{eq11}
\eea
$\sigma$ ($\tilde{\sigma}$) is the light (heavy) particle-nucleon
cross section and $\sigma_{abs}=(1-\epsilon)\tilde{\sigma}$ ($\epsilon$
can be interpreted as the probability for the heavy particle to survive
in one interaction,
$\epsilon=1$ ($\simeq 0$) for open charm (charmonium)).
Terms with $\sigma$
correspond to the modification (shadowing) of nucleon structure
functions inside nuclei; in Eq. (\ref{eq10}) it has been described by
an eikonal model using a multi-pomeron factorized vertex, although other
models could be used (as a sum of fan diagrams, i.e. the Schwimmer model,
see \cite{Bo93,SFN}).

The consequences of this approach, neglecting the modifications of
nucleon structure functions inside nuclei, are the following:
at high energies open charm and charmonium are e\-qual\-ly
absorbed (no $\sigma_{abs}$ appears in Eqs. (\ref{eq10}) and (\ref{eq11})),
see also
\cite{Bo93});
the low (\ref{eq7}) and high (\ref{eq11}) energy formulae
differ for charmonium up to 20 \% at the highest energies
(being the difference of order $(\tilde{\sigma}A/R_A^2)^3$,
$\tilde{\sigma}\simeq \sigma_{abs}$); and the exact \cite{Br98}
and probabilistic results differ $\sim 1\div 2 $ \% at $\sqrt{s}=20$ GeV.

These results can be interpreted as follows: At low energies only the first
interaction is effective for producing heavy flavor (so production is
proportional to $A$), subsequent
interactions can only absorb it. At high energies all interactions are
simultaneous, so the absorption mechanism of subsequent interactions is
no longer effective; instead the full multiple interaction formalism,
which suppresses equally both hidden and open heavy flavor production, has
to be considered.
Usual factorization (i.e. separation between partonic densities and
parton-parton cross sections, Eq. (\ref{eq1})) is broken and an
additional suppression factor is always present (except for
open heavy flavor production at low energies).
While at finite energies it is not
possible to separate in this additional factor
the modification of nucleon parton densities inside
nuclei from the scattering of the heavy partons,
this separation is
recovered at low
energies (Eq. (\ref{eq7}),
where there is no modification of nucleon parton densities)
and also, for fixed impact parameter $b$, at asymptotically high
energies (Eq. (\ref{eq10})).

\begin{figure}[hbt]
\label{fig10}
\caption{$A_{eff}$ versus $\protect{\sqrt{s}}$
for $x_F=0$ for charmonium and open
charm production in $pPb$ collisions: exact result
\protect{\cite{Br98}} (solid line), probabilistic formula (Eq. (7), dotted
line), asymptotic formula (Eq. (11), dashed line) and exact result with
nuclear modifications of structure functions as explained in the text
(dashed-dotted line).}
\epsfig{file=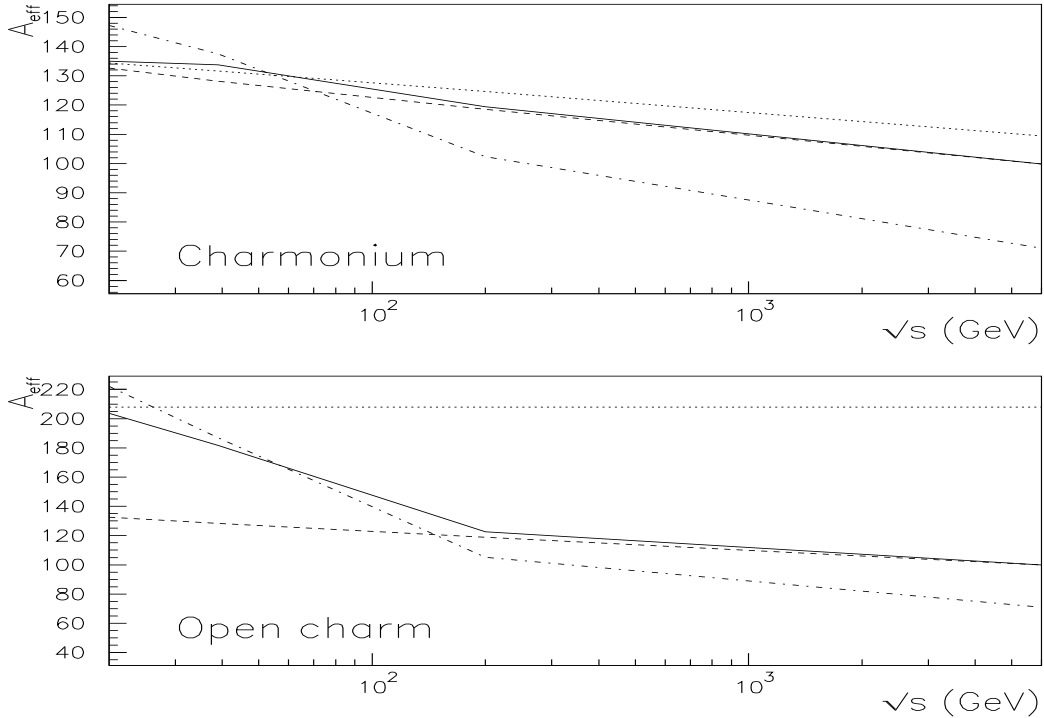,height=11cm,width=15.5cm,angle=0}
\end{figure}
\begin{figure}[hbt]
\label{fig11}
\caption{The same as Fig. 10 but for $x_F=0.5$.}
\epsfig{file=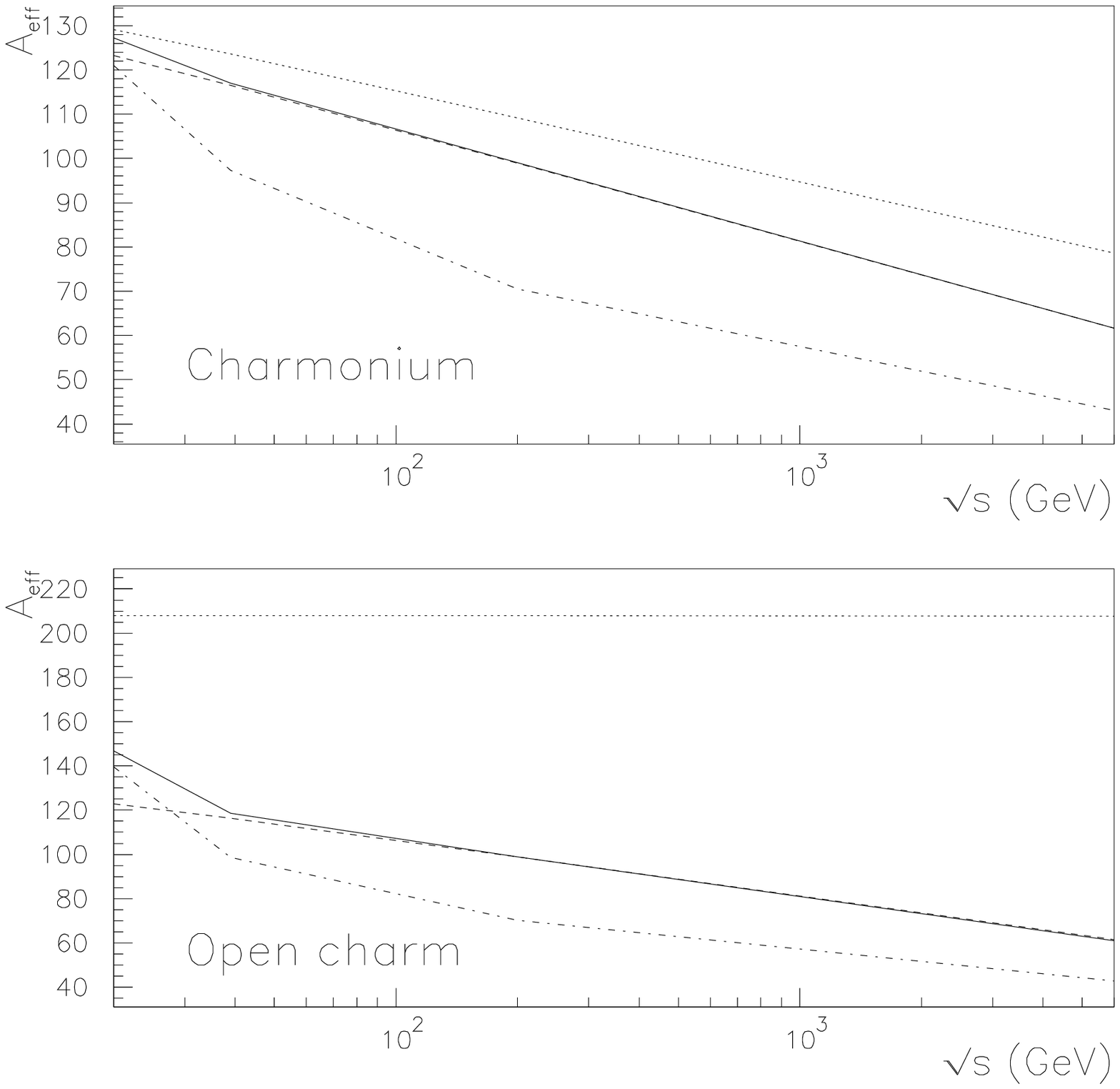,height=11cm,width=15.5cm,angle=0}
\end{figure}
Results for $pPb$ collisions \cite{Br98},
in the form of the variation with energy
of $A_{eff}=A^\alpha$, are presented in Figs. 10 and 11. We use
$\tilde{\sigma} \propto (xs)^{0.08}$
(normalized to 7 mb at $p_{lab}=$ 200 GeV/c \cite{ollitrault}),
$\epsilon=$ 1 (0.001) for open (hidden) charm and a standard
Woods-Saxon nuclear density. Just to give some estimation, the effect of
nuclear structure functions
has been taken into account considering it factorized and
computed following \cite{noso,pss}, see Section \ref{sec2}.
It can be observed that the exact result provides a smooth transition
between the low energy and the high energy regimes (the latter already
reached
at $\sqrt{s}=200$
GeV) and that the effect of the nuclear modification of structure
functions varies, at $x_F=0$,
from antishadowing at low energies to shadowing at high
energies.

Turning back to the $x_F$ dependence of charmonium nuclear absorption,
it is clear (see for example \cite{exp}) that the modification of the
nucleon parton distributions inside nuclei cannot reproduce the data:
for different energies,
there is no scaling of absorption in $x_b$ (corresponding to the target
nucleus), while there is approximate
scaling in $x_F=x_a-x_b$. Nevertheless nuclear
structure functions have to be taken into account. In Fig. 12
calculations \cite{pss} of $\alpha$ versus $x_F$ for $pW$ collisions
at $p_{lab}=800$ GeV/c
are presented, with the same parameters as in Subsection 2.2;
normalization is obtained using Eq. (\ref{eq7}) with $\sigma_{abs}\simeq
6$ mb (taken constant with $x_F$).
Clearly the modification of the nucleon parton distributions
inside nuclei accounts for part of the effect, except at the highest
$x_F$. More data, both at 800 GeV/c, as those from \cite{exp2}, and
at lower energies, have to be examined (following
e.g. the proposals of \cite{Br98,Bo93,Ko97}) in order to explain the
behavior of the absorption with $x_F$.

As a comment, let us stress that $\alpha$ is a misleading variable.
For example, in terms of $\alpha$ the experimental results of \cite{exp}
and \cite{exp2} seem to indicate some anomalous behavior at $x_F\simeq 0$
in $pA$ collisions at 800 GeV/c:
absorption increases going from $x_F\simeq 0.15$ to $x_F\simeq 0.65$
and
also from $x_F\simeq 0.15$ to $x_F\simeq 0$. If one
expresses these results in terms of the ratio $W/C$, which can be done
both for the E772 \cite{exp} and E789 \cite{exp2} data, this
ratio turns out to be monotonically decreasing with $x_F$ from
$x_F\simeq 0$ to 0.65 ($0.780\pm 0.074$, $0.776\pm 0.059$,
$0.746\pm 0.046$, $0.741\pm 0.020$,
$0.729\pm 0.034$, $0.649\pm 0.051$, $0.604\pm 0.084$, $0.571\pm 0.155$
for $x_F=-0.023$, 0.032, 0.16, 0.26, 0.36, 0.46, 0.55, 0.65
respectively, with only statistical errors taken into account).
So, it follows the trend that can be seen in Fig.
12 and no strange behavior can be deduced.

\section{Conclusions}\label{sec4}

Effects on heavy flavor and Drell-Yan \cite{noso} and
charmonium \cite{pss} production
of the modification of nucleon structure
functions inside nuclei have been examined.
They are relatively small at
low $x_F$, low $\sqrt{s}$ but become very important at high $x_F$ and
for RHIC and LHC energies. Thus, more experimental
and theoretical effort is needed to reduce the uncertainties in the
extrapolation to small $x$.
For charmonium nuclear absorption is,
at low energies and $x_F\sim 0$, of greater importance 
than nuclear effects on parton densities; for higher energies or
larger $x_F$ both
effects have to be taken into account.
\begin{figure}[hbt]
\caption{Comparison of $x_F$ dependence of nuclear corrections by
modifications of parton densities inside nuclei
with experimental data
of 800 GeV/c protons incident on a tungsten target from
\protect{\cite{exp}}. Theoretical calculations have been normalized to
one of the less shadowed experimental points (see text).
Lines follow the same convention as in Fig. 6.}
\epsfig{file=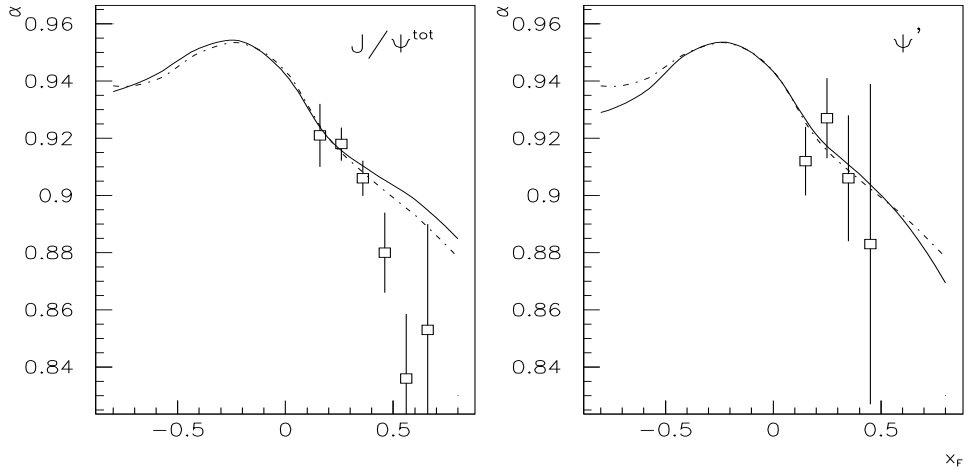,width=15.5cm,height=13cm,
angle=0,bbllx=0pt,bblly=0pt,
bburx=600pt,bbury=600pt}
\end{figure}

\vskip -5 cm
Formulae for nuclear production of open and hidden
heavy flavor have been presented \cite{Br98}, using a
relativistic Glauber-Gribov formalism in which the standard
probabilistic formula for charmonium absorption has been derived
as a low energy
limit of an exact expression valid at all energies.
The numerical accuracy of the probabilistic formula at available
energies has been checked.
A striking prediction of this approach \cite{Br98,Bo93} is that,
at high energies,
open charm will asymptotically be as suppressed as charmonium.

New data \cite{na50}
on $J/\psi$ suppression in $PbPb$ collisions
at SPS energies have produced great
excitement as a possible signal of new physics (i.e. Quark-Gluon Plasma).
In view of all the uncertainties commented in this contribution, detailed
tests of all our conventional ideas about open and hidden heavy flavor
production off nuclei are needed before any
quantitative statement on the existence of new physics
in heavy ion collisions can be made.

\vskip 0.5 cm
\noindent {\bf Acknowledgments:} The results presented in Section
\ref{sec2} have been obtained in collaboration with
C. Pajares and C. A. Salgado
(Santiago de Compostela) and Yu. M. Shabelski (Gatchina); those of
Section \ref{sec3}, in collaboration with M. A. Braun (St.
Petersburg), A. Capella (LPTHE Orsay), C. Pajares and C. A. Salgado.
I thank A. Capella
for comments on the manuscript and C.
A. Salgado for a careful reading, and C. Louren\c co, H. Satz, J.
Seixas and X.-N. Wang for their kind invitation to such a nice and
interesting workshop. I also thank S. J. Brodsky, M. Cacciari, K. J.
Eskola, E. G. Ferreiro,
R. V.
Gavai, A. B.
Kaidalov, D. Kharzeev, G. P. Korchemsky,
A. Krzywicki, M. Nardi, P. V.
Ruuskanen, H. Satz, R. Vogt and
X.-N. Wang for useful discussions, and C. Louren\c co and P. L. McGaughey
for discussions and
information on the experimental data. Financial support of Direcci\'on
General de Investigaci\'on Cient\'{\i}fica y T\'ecnica of Spain,
Instituto Superior T\'ecnico  of Lisbon and
Laborat\'orio de Instrumenta\c c$\tilde{\mbox{a}}$o
e F\'{\i}sica Experimental de Part\'{\i}culas 
of Lisbon is gratefully acknowledged.

\end{document}